\documentclass[pre,aps,twocolumn,superscriptaddress]{revtex4-1}
\pdfoutput=1
\usepackage{amsmath, amsthm, amssymb}
\usepackage{amsfonts}
\usepackage{graphicx}
\usepackage{dcolumn}
\usepackage{bm}
\usepackage{textcomp}
\usepackage[normalem]{ulem}

\usepackage[titletoc,title]{appendix}

\usepackage{epsfig}
\usepackage{subfigure}
\usepackage{url}
\usepackage{float}
\usepackage{cases}

\usepackage{colordvi}
\usepackage[usenames,dvipsnames]{xcolor}

\usepackage[colorlinks=true, urlcolor=blue, anchorcolor=blue, citecolor=blue,filecolor=blue,linkcolor=blue,menucolor=blue
]{hyperref}

\usepackage[titletoc,title]{appendix}

\usepackage{color}

\begin{document}
\title{Velocity fluctuations of stochastic reaction fronts propagating into an unstable state: strongly pushed fronts}

\author{Evgeniy Khain}
\email{khain@oakland.edu}
\affiliation{Department of Physics, Oakland University, Rochester, MI 48309, USA}
\author{Baruch Meerson}
\email{meerson@mail.huji.ac.il}
\affiliation{Racah Institute of Physics, Hebrew University of
Jerusalem, Jerusalem 91904, Israel}
\author{Pavel Sasorov}
\email{Pavel.Sasorov@eli-beams.eu}
\affiliation{Institute of Physics CAS, ELI Beamlines, 182 21 Prague, Czech Republic}
\affiliation{Keldysh Institute of Applied Mathematics, Moscow 125047, Russia}

\begin{abstract}
The empirical velocity of a reaction-diffusion front, propagating into an unstable state, fluctuates because of the shot noises of the reactions and diffusion. Under certain conditions these fluctuations can be described as a diffusion process in the reference frame moving with the average velocity of the front. Here we address pushed fronts, where the  front velocity in the deterministic limit  is affected by higher-order reactions and is therefore larger than the linear spread velocity.  For a subclass of these fronts -- strongly pushed fronts -- the effective diffusion constant $D_f\sim 1/N$ of the front can be calculated, in the leading order, via a perturbation theory in $1/N \ll 1$, where $N\gg 1$ is the typical number of particles in the transition region. This perturbation theory, however, overestimates the contribution of a few fast particles in the leading edge of the front. We suggest a more consistent calculation by introducing  a spatial integration cutoff at a distance beyond which the average number of particles is of order 1. This leads to a non-perturbative correction to $D_f$ which even becomes dominant close to the transition point between the strongly and weakly pushed fronts.  At the transition point we obtain a logarithmic correction to the $1/N$ scaling of $D_f$. We also uncover another, and quite surprising, effect of the fast particles in the leading edge of the front. Because of these particles, the position fluctuations of the front can be described as a diffusion process only on very long time intervals with a duration $\Delta t \gg \tau_N$, where $\tau_N$ scales as $N$. At intermediate times the position fluctuations of the front are anomalously large and  non-diffusive. Our extensive Monte-Carlo simulations of a particular reacting lattice gas model support these conclusions.

\end{abstract}

\maketitle

\nopagebreak

\section{Introduction}
\label{intro}

Effects of shot noise on the propagation of macroscopic reaction-diffusion fronts have attracted much attention in physics, chemistry and biology \cite{vanSaarloos03,Panja}. The shot noise is a natural consequence of the discreteness of the constituent particles and of the randomness of elemental processes of reactions and diffusion. The shot noise causes a systematic shift in the mean front position, compared with the deterministic prediction,  and position fluctuations around the mean. Typical fluctuations of the front position around the mean  can be usually described in terms of  front diffusion \cite{Panja}. A natural question then is how the corresponding diffusion constant $D_f$ scales with the characteristic number $N\gg 1$ of particles in the transition region of the front.  The answer to this question strongly depends on whether the front propagates into an unstable or a metastable (that is, linearly stable but nonlinearly unstable) state of the underlying deterministic theory \cite{vanSaarloos03,Panja}.  For fronts  propagating into a metastable state, $D_f$ exhibits the \textit{a priori} expected $1/N$ scaling \cite{Panja,MSK,KM}, and it can be calculated by using a stochastic reaction-diffusion equation, that governs the system, in conjunction with a perturbation expansion in $1/N\ll 1$ \cite{MSK}. At the other extreme one finds \emph{pulled} fronts. For this subclass of fronts, propagating into an unstable state, the asymptotic front velocity, as predicted by the underlying deterministic theory, is determined by the leading edge of the front, and it is equal to the linear spread velocity \cite{vanSaarloos03}.  The pulled fronts are extremely sensitive to the shot noise in their leading edge \cite{vanSaarloos03,Panja,Tsimring,Derrida1,Derrida06,MSfisher,MSV}. For such fronts the front diffusion constant $D_f$ scales as $\ln^{-3} N$ \cite{Derrida06}, that is it is rather large.

There is, however, an intermediate class of fronts that has received much less attention. These are \emph{pushed} fronts  propagating into an unstable state \cite{vanSaarloos03}. Their asymptotic velocity, as predicted by the deterministic theory, is affected by the higher-order reactions and, as a result, exceeds the linear spread velocity \cite{vanSaarloos03}. The effects of shot noise on the propagation of such fronts have been recently studied by Birzu \textit{et al.}~\cite{Birzu2018}. They described pushed fronts by a simplified stochastic partial differential equation (sPDE) which accounts for the shot noise of the particle reactions, but not of the particle diffusion.  They observed that the pushed fronts can be divided into two subclasses -- which we will call strongly pushed and weakly pushed -- in their relation to the perturbation theory  of Ref. \cite{MSK}. For the strongly pushed fronts the spatial integrals, entering the perturbative expression for $D_f$  in Ref.~\cite{MSK}, are convergent, and one obtains the same scaling behavior $D_f\sim 1/N$ as in the well-studied metastable case~\cite{Birzu2018}. For the weakly pushed fronts the perturbative expression  for $D_f$  in Ref.~\cite{MSK} is divergent, signalling a much larger, non-perturbative contribution to $D_f$ coming from the leading edge of the front \cite{Birzu2018}.

In this paper we deal with fluctuations of the strongly pushed fronts,  leaving the weakly pushed fronts for a future work.  As in our previous works on fluctuating reaction-diffusion fronts \cite{MSfisher,MSK,MSV,KM}, we consider a  class of reacting lattice gas models which involve random walk on a one-dimensional lattice and on-site reactions among a single species of particles. Our analysis applies, in a proper parameter region, to many sets of reactions for which the continuous-in-space deterministic limit of the model -- a deterministic reaction-diffusion equation -- describes a pushed front propagating into an unstable state. Upon a proper rescaling, this equation has a single dimensionless parameter of order unity which we call $\gamma$. This parameter affects the asymptotic propagation velocity of the front and determines whether the front is strongly or weakly pushed.  Taking into the account the shot noises of particle reactions and diffusion, one can obtain an sPDE \cite{MSK}, where the noise terms are small when $N\gg 1$.   For the strongly pushed fronts, which we focus on here, the perturbative expression  \cite{MSK} for $D_f$ turns out to be convergent, as in the simplified model of Ref. \cite{Birzu2018}. However, by extending  the spatial integration in that perturbative expression into the region where the average number of particles drops below $O(1)$, this calculation significantly overestimates $D_f$. To begin with, the sPDE \cite{MSK} is not expected to be correct there. But even if it were correct, this region is dominated by noise and,  regardless of $N$,  there is no reason to believe that the perturbation theory \cite{MSK} remains applicable.

To obtain a better approximation for $D_f$, we introduce an integration cutoff  at a distance beyond which the average number of particles is of order 1. The cutoff leads to a negative non-perturbative correction to $D_f$ which scales as $N^{-1-\nu}$, where the parameter $0<\nu<1$ depends only on $\gamma$.
The non-perturbative correction is much larger than the expected $1/N^2$ correction that would come from the second and third orders of the perturbation theory of Ref. \cite{MSK}. Even more importantly, it becomes dominant close to  the transition point between the strongly pushed and weakly pushed fronts.  At the transition point we obtain a logarithmic correction to the $1/N$ scaling: $D_f\sim \ln N/N$.

We also find a striking additional effect of the few fast particles at the leading edge of the front, and this effect is exclusive to pushed fronts.  As we show here,  the position fluctuations of the strongly pushed fronts can be described as
diffusion in the moving frame only when they are observed over very long time intervals, $\Delta t \gg \tau_N $, where
the characteristic time $\tau_N$ scales as $N$.  At intermediate times the front position fluctuations in the moving frame are non-diffusive. Remarkably, they are almost independent of $N$ and therefore very large.

Our theoretical results are supported by extensive Monte-Carlo simulations, which we performed in the region of parameters where
the front is (i) strongly pushed, and (ii) relatively close to the  transition point between the strongly and weakly pushed fronts.

The remainder of the paper is organized as follows. In Sec. \ref{theory} we formulate the model, present the governing equations and evaluate the diffusion coefficient $D_f$ of the strongly pushed fronts.  In Sec. \ref{transient} we discuss the large deviation function of the front velocity fluctuations and the fluctuations of the empiric velocity at intermediate times.  Section \ref{simulations} presents our simulation method and a comparison of simulation results with theory. Section \ref{summary} contains a brief
summary and discussion of our results.

\section{Theory}
\label{theory}

\subsection{Model and governing equations}

\label{general}

\begin{figure}[ht]
\includegraphics[width=0.38\textwidth,clip=]{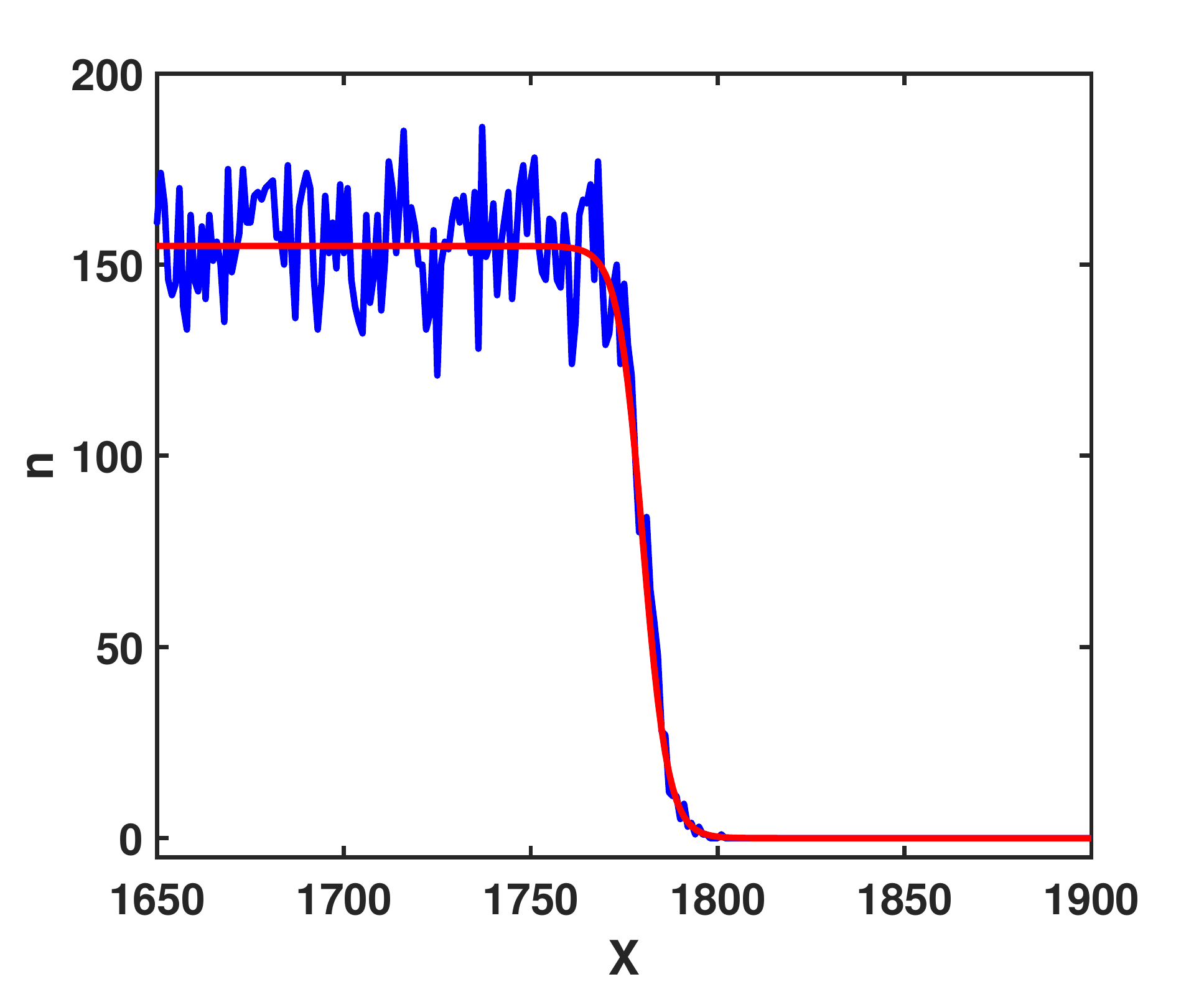}
\caption{The density profile of a propagating stochastic pushed front on the lattice with lattice constant $h=1$. Shown is the simulated number density of the particles (the number of particles per site) as a function of the integer coordinate $X=i$. The smooth curve is the theoretical deterministic profile~(\ref{ansol}), shifted to match the simulated front. The parameters are: $\alpha=1$, $\beta = 2/45$, $\sigma = 1/900$ and $D_0 = 50$.  For these parameters $K=60$, $\gamma = 1.5$ and $N= 300$.}
\label{fig:frontprofile2}
\end{figure}

The departure point of our analysis is a microscopic model which involves a single species of particles, which we call $A$, residing on a one-dimensional lattice with lattice constant $h$. The number of particles $n_i$ on each lattice site $i$ varies in time as a result of two types of Markov stochastic processes: on-site reactions among the particles, and independent random walk (where a particle hops with equal probabilities to \emph{any} of the two adjacent sites) with the rate constant $D_0$. A simple and generic example, that we will be mostly working with, includes three on-site reactions:  the branching $A\to 2A$ with the rate constant $\alpha$, and the reactions $2A\to 3A$ with the rate constant $\beta$ and $3A\to 2A$ with the rate constant $\sigma$. We will assume a strong rate disparity, $D_0\gg \alpha\gg \beta \gg \sigma$, and introduce two dimensionless parameters $K=3\beta/(2\sigma)\gg 1$ and  $\gamma= 8\alpha\sigma/(3 \beta^2)$, which we will assume to be $O(1)$ \cite{disparity}.  The characteristic steady-state population size on a single site scales as $K$. The parameter $\gamma$ determines the front type, as we will see shortly. A snapshot of  such a stochastic front is shown in Fig.~\ref{fig:frontprofile2}. We obtained it in Monte Carlo simulations, described in Sec.~\ref{simulations}.

As $D_0$ is very large compared with $\alpha$,  the random walk on the lattice can be approximated by the continuous Brownian motion on the line, with the diffusion constant $D=D_0h^2/2$. In this regime, and for typical (not large) fluctuations around the deterministic front, one can derive a continuous sPDE for the rescaled particle density $u(x,t)=n_i/K$ as a function of rescaled time $t$ and coordinate $x$  \cite{MSK,MS}:
\begin{eqnarray}
\partial_t u(x,t) &=& \alpha f(u)+ D \partial_x^2 u+\sqrt{\frac{\alpha g(u)\,h}{K}}\, \eta(x,t) \nonumber \\
&+&\partial_x \left[\sqrt{\frac{2 u(x,t)\, Dh}{K}}\, \chi(x,t)\right]\,,
\label{rw040}
\end{eqnarray}
where $\eta(x,t)$ and $\chi(x,t)$ are independent Gaussian noises  which are $\delta$-correlated  both in space and in time with zero mean and unit magnitude:
\begin{equation}
\left\langle \eta(x,t)\eta(x^{\prime},t^{\prime})\right\rangle = \left\langle \chi(x,t)\chi(x^{\prime},t^{\prime})\right\rangle = \delta(x-x^{\prime})\, \delta(t-t^{\prime}).
\label{rw060}
\end{equation}
Further, for the three reactions that we have introduced,
\begin{eqnarray}
f(u)&=& u+\frac{2u^2}{\gamma}-\frac{u^3}{\gamma}\,,\label{f}\\
g(u)&=& u+\frac{2u^2}{\gamma}+\frac{u^3}{\gamma}\,.\label{g}
\end{eqnarray}
The terms with $\eta(x,t)$ and $\chi(x,t)$ describe the shot noises of the particle reactions and of the particle diffusion, respectively. In Ref.~\cite{Birzu2018} the noise of the random walk  was not taken into account. This omission is not essential, except for numerical factors, in the description of the effective diffusion of the front. The omission becomes, however, crucial in the description of the positive fluctuations of the front velocity, coming from a few particles outrunning the front.  As we show here, at intermediate times these fluctuations become very important, as they determine \emph{typical} fluctuations of the front velocity. In order to describe them, however, we will have to abandon the sPDE (\ref{rw060}) altogether and return to the microscopic model, see Sec.~\ref{transient}.

It is convenient to work in rescaled coordinates. Let us measure time in units of $1/\alpha$ and the coordinate $x$ in units of the diffusion length  $\sqrt{D/\alpha}$. As a result, the units of velocity are $\sqrt{\alpha D}$, and the units of diffusion constants are $D$.  The rescaled form of Eq.~(\ref{rw060}) is
\begin{equation}
 \label{2}
  \partial_{t} u(x,t) = f(u) +  \partial_{x}^2 u(x,t) + \frac{1}{\sqrt{N}} \,R(x,t,u) \,,
\end{equation}
where
\begin{equation}
R(x,t,u)= \sqrt{g(u)}\, \eta(x,t)+\partial_{x}[\sqrt{2u}\, \chi(x,t)]\,, \label{R}
\end{equation}
and $N\gg 1$, the characteristic number of particles in the transition region of the front,  is formally defined as $N=K\sqrt{D/\alpha h^2}\equiv K\sqrt{D_0/2\alpha}$. The rescaled noises $\eta(x,t)$ and $\chi(x,t)$ obey Eq.~(\ref{rw060}), but with the rescaled coordinate and time.

In the absence of the noise terms,  Eq.~(\ref{2}) is a well-known deterministic reaction-diffusion equation
\begin{eqnarray}
 \partial_t u &=& f(u)+\partial_x^2 u\,.
 \label{deteq}
\end{eqnarray}
For our particular set of reactions the polynomial  $f(u)$ has two nonnegative roots: a stable one,
\begin{equation}\label{stableroot}
u=U_0=1+\sqrt{1+\gamma}\,,
\end{equation}
and an unstable one, $u=0$. Suppose that the boundary conditions are $u(x\to -\infty, t)=U_0$ and $u(x\to \infty, t)=0$. Equation~(\ref{deteq})
has a traveling-front solution (TFS) $u(x,t)=U(\xi)$, $\xi=x-c_0 t$, which obeys the ordinary differential equation (ODE)
\begin{equation}\label{PFmfs}
    U^{\prime\prime}+c_0 U^{\prime} +U+\frac{2U^2}{\gamma}-\frac{U^3}{\gamma} =0\,.
\end{equation}
The solution of this ODE, subject to the boundary conditions, is unique up to translations in $\xi$, and quite simple:
\begin{equation}
\label{ansol}
U(\xi)=\frac{U_0}{1+e^{\lambda\xi}} \,.
\end{equation}
The spatial decay rate of the front, $\lambda$, is equal to
\begin{equation}\label{lambda}
\lambda=\frac{\sqrt{\gamma +1}+1}{\sqrt{2\gamma }} \,,
\end{equation}
and the (deterministic) front velocity is
\begin{equation}\label{PFc0}
    c_0=\frac{3 \sqrt{\gamma +1}-1}{\sqrt{2 \gamma}}>0\,.
\end{equation}
Importantly, Eqs.~(\ref{ansol})-(\ref{PFc0})  correctly describe an asymptotic front, developing from a localized initial condition for Eq.~(\ref{deteq}), only when the front is pushed, that is $c_0>2$, see \textit{e.g.} Ref. \cite{vanSaarloos03}. This occurs at $\lambda>1$ or, in terms of $\gamma$, at $0<\gamma<8$. For $\lambda<1$, or $\gamma>8$, the front is pulled: here Eqs.~(\ref{ansol})-(\ref{PFc0}) are inapplicable, and the correct asymptotic front velocity $c_0$ is not affected by the nonlinear terms of the function $f(u)$. Rather, it coincides with the linear spread velocity for Eq.~(\ref{deteq}) which, for $f(u)$ from Eq.~(\ref{f}), is equal to $2$. The present paper deals only with the pushed fronts.
Figure~\ref{fig:frontprofile2} shows a comparison of Eq.~(\ref{ansol}) with a snapshot of simulated stochastic front.

\subsection{Diffusion of strongly pushed stochastic fronts}
\label{diffusion}

Now let us return to the sPDE (\ref{2}).    Using the small parameter  $N^{-1/2} \ll 1$, the authors of Ref. \cite{MSK} (see also earlier works \cite{Mikhailov2,Pasquale,Rocco} on qualitatively similar front models) developed a perturbation theory for a model similar to Eq.~(\ref{2}), but where the front propagates into a metastable state. In the first order in  $N^{-1/2} \ll 1$ one obtains a closed-form analytic result for the effective diffusion constant $D_f$, which describes typical fluctuations of the front position around its mean \cite{MSK}:
\begin{equation}\label{DF010}
D_f=\frac{A_{\infty}}{N}\, ,
\end{equation}
where
\begin{equation}\label{DF020}
A_{\infty}=\frac{J_1(\infty) + J_2(\infty)}{J_3^2(\infty)}\, ,
\end{equation}
and the functions $J_1(\xi)$,  $J_2(\xi)$, and  $J_3(\xi)$ are given by the following expressions:
\begin{eqnarray}
  J_1(\xi)&=&\frac{1}{2}\int\limits_{-\infty}^{\xi} g\left[U(\xi)\right]\left[U^\prime(\xi)e^{c_0\xi}\right]^2 \, d\xi\, , \label{DF030}\\
  J_2(\xi)&=&\int\limits_{-\infty}^{\xi} U(\xi)\left[\left(U^\prime(\xi)e^{c_0\xi}\right)^\prime\right]^2 \, d\xi\, , \label{DF040} \\
 J_3(\xi)&=&\int\limits_{-\infty}^{\xi} \left[U^\prime(\xi)\right]^2e^{c_0\xi} \, d\xi\, .\label{DF050}
\end{eqnarray}
For a front propagating into a metastable state, all three quantities $J_1(\infty)$, $J_2(\infty)$ and $J_3(\infty)$  are finite, and the perturbation theory \cite{MSK} is consistent.  When applying Eqs.~(\ref{DF020})-(\ref{DF050}) to the pushed fronts propagating into an unstable state, one can see that the quantity $J_3(\infty)$ is finite for all pushed fronts \cite{Birzu2018}, that is for $0<\gamma<8$.  The quantities $J_1(\infty)$ and $J_2(\infty)$, however, are finite only for $c_0>3/\sqrt{2}$  or $0<\gamma<16/9$: these are the strongly pushed fronts. For $c_0<3/\sqrt{2}$, or $\gamma>16/9$ (the weakly pushed fronts)  the quantities $J_1(\infty)$ and $J_2(\infty)$ are infinite, leading to a divergent expression for $D_f$ \cite{Birzu2018}. This formal divergence signals a much larger, non-perturbative contribution to $D_f$ for the weakly pushed fronts, coming from the leading edge of the front.

For the strongly pushed fronts, a convergent expression for $D_f$  gives a valid leading-order behavior of $D_f$ at $N\to\infty$.  However, extending the spatial integration into the leading-edge region, where there are only a few particles, or no particles at all, one overestimates $D_f$. To obtain a more accurate expression for $D_f$, we will estimate a non-perturbative negative correction to the leading-order expression, described by Eqs.~(\ref{DF010}) and (\ref{DF020}). We will achieve it by introducing an integration cutoff in Eqs.~(\ref{DF030})-(\ref{DF050}) at a distance beyond which the average number of particles is $O(1)$. As we will see shortly, this non-perturbative correction for the strongly pushed fronts  scales with $N$ as $N^{-1-\nu}$, where $0<\nu<1$ depends only on $\gamma$. The correction is much larger than the expected $O(N^{-2})$ correction that would come from the perturbation theory \cite{MSK} in $1/N$, extended to second and third orders.
The $N^{-1-\nu}$ correction becomes quite significant in comparison with the leading order term $N^{-1}$  even for large $N$, when we approach the transition point between the strongly and weakly pushed fronts,  where $\nu$ tends to zero.

\begin{figure}[h]
\includegraphics[width=0.35\textwidth,clip=]{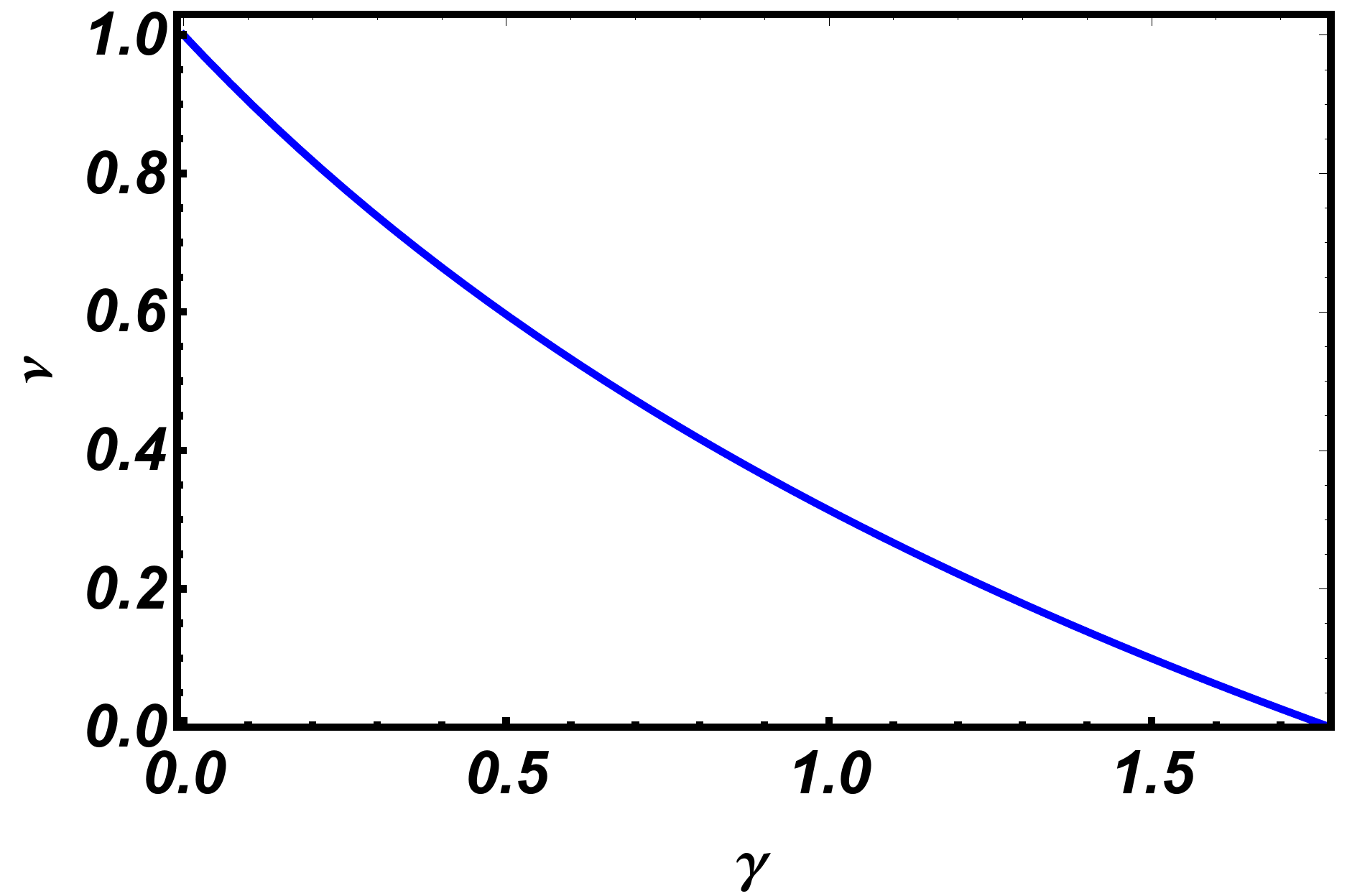}
\includegraphics[width=0.35\textwidth,clip=]{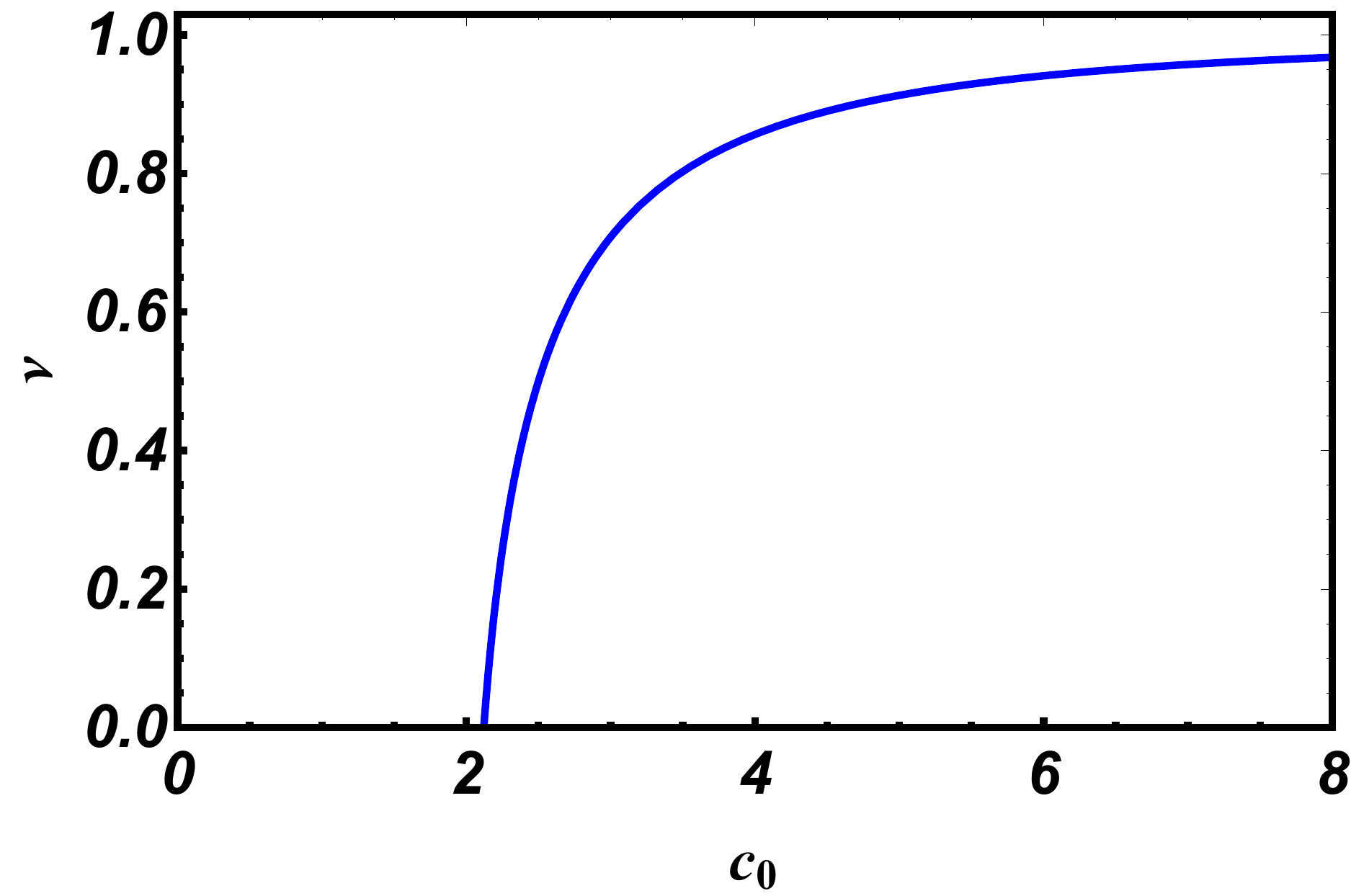}
\caption{$\nu$ versus $\gamma$ (top) and versus $c_0$ (bottom), as described by
Eqs.~(\ref{nugamma}) and~(\ref{DF126}), respectively.}
\label{nuvsgc}
\end{figure}

Implementing this program, we truncate the  integrals in Eqs.~(\ref{DF020})-(\ref{DF050})
at the point $\xi_0$ where
the average rescaled particle density $U(\xi)$ is equal to $1/(kN)$, where $k=O(1)$. We obtain
\begin{equation}\label{DF070}
D_f = \frac{A(N)}{N}\,, \quad \text{where}\quad A(N)=\frac{J_1(\xi_0) + J_2(\xi_0)}{J_3^2(\xi_0)}\,,
\end{equation}
$\xi_0$ is the root of the algebraic equation
\begin{equation}\label{DF080}
U(\xi_0)=\frac{1}{kN}\,,
\end{equation}
and $N\gg 1$. We can write
\begin{equation}\label{DF090}
J_{i=1,2,3}(\xi_0)=J_i(\infty) - \int\limits_{\xi_0}^\infty \, \dots\, d\xi\, .
\end{equation}
As $\xi_0\gg 1$, the integrals from $\xi_0$ to $\infty$ can be evaluated by using the leading-edge asymptotic of $U(\xi)$ from Eq.~(\ref{ansol}):
\begin{equation}\label{Ufar}
U(\xi)\simeq U_0 e^{-\lambda\xi} \,.
\end{equation}
In addition, we can set there $g(U)\simeq U$. As a result,
\begin{eqnarray}
\!\!\!\!\!J_1(\xi_0)\!&\simeq&\!J_1(\infty)-\frac{U_0^3\lambda^2}{2(3\lambda-2c_0)} e^{-(3\lambda-2c_0)\xi_0}\nonumber \\
\!&\simeq&\!J_1(\infty)-\frac{U_0^3\lambda^2}{2(3\lambda-2c_0)} (U_0kN)^{-(3\lambda-2c_0)/\lambda},
\label{DF100}\\
\!\!\!\!\!J_2(\xi_0)\!&\simeq&\!J_2(\infty)-\frac{U_0^3 \lambda^2 (\lambda-c_0)^2}{3\lambda-2c_0} e^{-(3\lambda-2c_0)\xi_0} \nonumber \\
\!&\simeq&\! J_2(\infty)-\frac{U_0^3 \lambda^2 (\lambda-c_0)^2}{3\lambda-2c_0}  (U_0kN)^{-(3\lambda-2c_0)/\lambda},
\label{DF110}\\
\!\!\!\!\!J_3(\xi_0)\!&\simeq&\!J_3(\infty)-\frac{U_0^2\lambda^2}{2\lambda-c_0} e^{-(2\lambda-c_0)\xi_0},
\label{DF120}
\end{eqnarray}
where $\lambda$ and $c_0$ are given by Eqs.~(\ref{lambda}) and~(\ref{PFc0}), respectively, and
\begin{equation}\label{xi0}
\xi_0 \simeq \frac{1}{\lambda}\,\ln(U_0kN)\,,
\end{equation}
as obtained from Eq.~(\ref{DF080}).  Since $\gamma>0$, the factor $2\lambda-c_0$ in the exponent of the subleading term in Eq.~(\ref{DF120})  is larger than the factor $3\lambda -2c_0$ in the exponents of the subleading terms in Eqs.~(\ref{DF100}) and (\ref{DF110}). Therefore, the subleading term in Eq.~(\ref{DF120}) should be neglected  to avoid excess of accuracy, and we obtain
\begin{eqnarray}\label{DF124}
A(N) &\simeq & A_{\infty}-\frac{ U_0^3 \lambda \left[1+2(\lambda-c_0)^2\right]}{2 \nu J_3^2(\infty)}
 (U_0kN)^{-\nu}\, ,
\end{eqnarray}
where
\begin{equation}\label{nugamma}
\nu = 3-\frac{2 c_0}{\lambda} = \frac{ 8 \sqrt{\gamma+1}-3 \gamma-8}{\gamma}\,, \quad 0<\gamma<\frac{16}{9}.
\end{equation}
As $\gamma$ increases from $0$ to $16/9$, $\nu$ decreases from $1$ to $0$, see the top panel of Fig.~\ref{nuvsgc}.

Importantly,  the following relationship between $\lambda$ and $c_0$,
\begin{equation}\label{DF126}
\lambda=\frac{c_0+\sqrt{c_0^2-4}}{2}\,
\end{equation}
holds for all pushed fronts. This formula is independent of the particular three-reaction model. Indeed, it is obtained by solving the following generic equation:
\begin{equation}\label{linearizedODE}
U^{\prime\prime}+c_0 U^{\prime} +U =0
\end{equation}
[a linearized version of  Eq.~(\ref{PFmfs})], and choosing, among the two exponential solutions, the one with the \emph{higher} decay rate \cite{vanSaarloos03}. Using Eq.~(\ref{DF126}), we can express the exponent $\nu$ through $c_0$ in a model-independent form:
\begin{equation}\label{DF128}
\nu=c_0\sqrt{c_0^2-4}-c_0^2+3\,,
\end{equation}
see the bottom panel of Fig.~\ref{nuvsgc}. The other coefficients in Eq.~(\ref{DF124}) -- $A_{\infty}, U_0$ and $J_3(\infty)$ -- are model-dependent. The condition $\nu>0$ guarantees that the correction $O(N^{-\nu})$ in Eq.~(\ref{DF124}) is small at large $N$. This condition is equivalent to $c_0>3/\sqrt{2} =2.12132 \dots$; it defines the strongly pushed fronts. In our particular three-reaction model it reads $\gamma < 16/9$.

The exponent $\nu$ becomes small as one approaches the strong-weak transition point $c_0=2.12132 \dots$, or $\gamma=16/9$. Therefore, at not too large values of $N$, the correction $\sim N^{-\nu}$ in Eq.~(\ref{DF124}) (and the corresponding correction $\sim N^{-1-\nu}$ to $D_f$)  can be quite significant.

\subsubsection{$\gamma=3/2$, or $c_0=2.1612\dots$}

As a particular example, let us evaluate $A(N)$ from Eq.~(\ref{DF124}) for $\gamma=3/2$. Here $\lambda\simeq 1.4902$, $U_0\simeq 2.5811$, and $\nu\simeq  0.09941$.  A numerical evaluation of the integrals $J_{1,2,3}$ yields $J_1(\infty)\simeq 114.66$, $J_2(\infty)\simeq 118.86$ and $J_3(\infty) \simeq 8.42097$. As a result, $A_{\infty}\simeq 3.2931$, and  Eq.~(\ref{DF124}) yields
\begin{equation}\label{DF130}
A(N)\simeq 3.2931\,- 3.1438(k N)^{-0.09941}\, .
\end{equation}
The exponent $\nu\simeq 0.09941$ is very small, so the correction is quite significant. In this approximate theory  the parameter $k=O(1)$ is arbitrary, and it can serve as an adjustable parameter. In Sec.~\ref{simvstheory} we will compare the prediction of (\ref{DF130}) with Monte-Carlo simulations of the
three-reaction system on the lattice.

\subsubsection{Approaching the strong-weak transition}

All the three integrals $J_{1,2,3}(\infty)$ depend on $\gamma$. In particular, the integrals  $J_{1,2}(\infty)$ diverge at infinity at the transition point $\gamma= 16/9$. Let us approximately evaluate $A(N)$ and, therefore, the diffusion constant $D_f$ of the front, in the vicinity of the transition point, and see the effect of cutoff there.

As $\gamma$ approaches $16/9$ from below, the integrals $J_{1,2}(\infty)$  are dominated by the region $\xi\gg 1$, where we can again use the large-$\xi$ asymptotic of $U(\xi)$ from Eq.~(\ref{Ufar}). As a result,  we find that $J_{1,2}(\infty)$ behave as
\begin{eqnarray}
J_1(\infty)&=&\left.\frac{U_0^3\lambda^2}{2}\right|_{\gamma=16/9}\,\,\frac{1}{3\lambda-2c_0}+\dots
\, ,
\label{DF140}\\
J_2(\infty)&=&U_0^3 \lambda^2 (\lambda-c_0)^2\big|_{\gamma=16/9}\,\,\frac{1}{3\lambda-2c_0}+\dots \, .
\label{DF150}
\end{eqnarray}
As $\gamma \to 16/9$, the denominator $3\lambda-2c_0$ vanishes, and $J_{1,2}(\infty)$ diverge. The dots $\dots$ in Eqs.~(\ref{DF140}) and~(\ref{DF150}) denote subleading constant terms. These can be neglected, because their contribution to $D_f$ scales as $1/N$, which is small compared with the correction $O(N^{-\nu})$, coming from the integration from $\xi_0$ to $\infty$.

In contrast to $J_{1,2}(\infty)$, the  integral $J_{3}(\infty)$ is well-behaved at $\gamma=16/9$. Moreover, using Eq.~(\ref{DF120}), we can evaluate it exactly:
\begin{eqnarray}
\label{J3169}
J_3(\infty)\big|_{\gamma=16/9}&=& \frac{8}{9}\int_{-\infty}^{\infty} e^{\frac{3 \xi}{\sqrt{2}}}\,
   \text{sech}^4\left(\frac{\xi}{\sqrt
   {2}}\right)\,d\xi \nonumber \\
&=& \frac{20 \sqrt{2} \pi }{9} = 9.87307\dots\,.
\end{eqnarray}
The other parameters at the transition point are
$c_0=3/\sqrt{2}$, $\lambda=\sqrt{2}$, and $U_0=8/3$.  The evaluation of the subleading correction term is straightforward, and
we obtain:
\begin{eqnarray}\label{DF160}
A(N)&=&
\frac{\lambda\left[1+2(\lambda-c_0)^2\right]U_0^3}{2 J_3^2(\infty)}
\Big|_{\gamma=\frac{16}{9}}\,\,
\frac{1- (U_0kN)^{-\nu}}{\nu}+\dots \nonumber \\
&=& \frac{48\sqrt{2}}{25 \pi^2}\,\frac{1- (U_0kN)^{-\nu}}{\nu} +\dots \nonumber \\
 &=& 0.27511\dots\, \frac{1- (U_0kN)^{-\nu}}{\nu}+\dots \,,
\end{eqnarray}
where one can put $\nu = 2 \sqrt{2} \left(c_0-3/\sqrt{2} \right)$.
Since
$$
\lim_{\nu\to 0} \frac{1- (U_0kN)^{-\nu}}{\nu} = \ln (U_0kN)\,,
$$
we obtain the following asymptotic behavior of $A(N)$, and hence of $D_f(N)$, exactly at the strong-weak transition point:
\begin{equation}\label{DF170}
D_f(N)\big|_{\gamma=16/9}\simeq \frac{48\sqrt{2}}{25 \pi^2}\,\frac{\ln[(8/3)kN]}{N}\, .
\end{equation}
The numerical coefficient in Eq.~(\ref{DF170}) is model-dependent. However, the $\ln N/N$ scaling of $D_f$ at the transition between the strongly pushed and weakly pushed fronts is universal.

Going back to Eq.~(\ref{DF160}), we notice that, at $|\nu|\ll 1$, $A(N)$  is an analytic function of $\nu$.
This suggests that Eq.~(\ref{DF160}) is also correct for small negative $\nu$, that is for the \emph{weakly-pushed} fronts close to the transition point.

\section{Fast particles, large fluctuations and long transient}
\label{transient}

We have argued in the previous section that a few particles in the leading edge of the front introduce an important correction to the front diffusion constant. These particles also play an additional, and a truly dramatic, role at intermediate times, $1\ll \Delta t \ll N$.  In order to see it, let us introduce the probability distribution $P(c, \Delta t, N)$
of observing a specified empirical velocity of the front  $c=\Delta X/\Delta t$, on the time interval $\Delta t$. Here $\Delta X = X_f-X_0$ is the displacement of the front during this time interval. Let us denote the average value of the empirical velocity (which is the typical velocity of the front) by $\bar{c}$; it differs from $c_0$ by a correction which vanishes as $N\to \infty$. At $\Delta t\gg 1$, the probability of observing any value of $c$ different from the average value $\bar{c}$ is exponentially small, and $P(c, \Delta t, N)$ exhibits a large-deviation behavior
\begin{equation}\label{lnP}
- \ln P(c, \Delta t, N) \simeq \Delta t \,r(c,N)\,.
\end{equation}
with a rate function $r(c,N)$. This rate function has been unknown except its asymptotic at $|c-\bar{c}| \ll \bar{c}$. This asymptotic corresponds to the typical fluctuations of the front, which are describable by the perturbation theory in $1/N$, see  subsection \ref{diffusion}. In this regime,  and not too close to the strong-weak transition, $r(c,N)$ is a parabolic function of $c$:
\begin{equation}\label{rgauss}
r(c,N) \simeq \frac{\left(c-\bar{c}\right)^2}{4 D_f}  = \frac{N}{4 A(N)} \left(c-\bar{c}\right)^2\,.
\end{equation}
The function $A(N)$ is given by Eq.~(\ref{DF124}) but, for the purpose of this section, it will suffice to neglect the cutoff-induced correction and set $A(N)\simeq A_{\infty}=\text{const}$. Notice that, as $N\gg 1$, the parabola~(\ref{rgauss}) is very steep.

The large-deviation regime of \emph{positive} velocity fluctuations, $c-\bar{c} \gg \bar{c}$, is dominated by a very few particles, outrunning the front. Here the sPDE~(\ref{2}) is inapplicable, and we should return to the microscopic model. Because of the disparity of rates, we can account only for the random walk and the branching reaction $A\to 2A$ in this region of space, and neglect the higher-order reactions  $2A \to 3A$ and $3A\to 2A$. Since the hopping rate $D_0$ is very large, the branching random walk is equivalent to the branching Brownian motion (BBM). In the rescaled variables the branching rate and the diffusion constant of the particles are both equal to 1. The large-deviation properties of the BBM are quite well known \cite{Rouault,DMS}. The probability density of the empirical velocity has the form (\ref{lnP}) with the rate function
\begin{equation}\label{rBBM}
r(c,N) \simeq \frac{c^2}{4}-1\,,\quad c-\bar{c}\gg 1\,.
\end{equation}
To remind the reader: at large $N$  $\bar{c}$ is close to $c_0$ and larger than $2$ for the pushed fronts that we are studying. We emphasize that the parabola~(\ref{rBBM}) is independent of $N$.

In the large-deviation regime of \emph{negative} velocity fluctuations, $\bar{c}-c\gg \bar{c}$, the physics is very different. To significantly decelerate the front, the shot noise has to modify the density profile in the whole transition
region. Similar situations were previously studied for the fronts propagating into a metastable state \cite{MSK,KM}, and for the pulled fronts \cite{MSfisher,MSV}. Here one can apply the optimal fluctuation method (other names: WKB approximation, instanton method, \textit{etc.}) \cite{MSK,MSfisher,MSV}. For our present purposes it suffices to know that the rate function $r(c,N)$ is proportional to $N$ in this regime and very steep.

\begin{figure}[ht]
\includegraphics[width=0.35\textwidth,clip=]{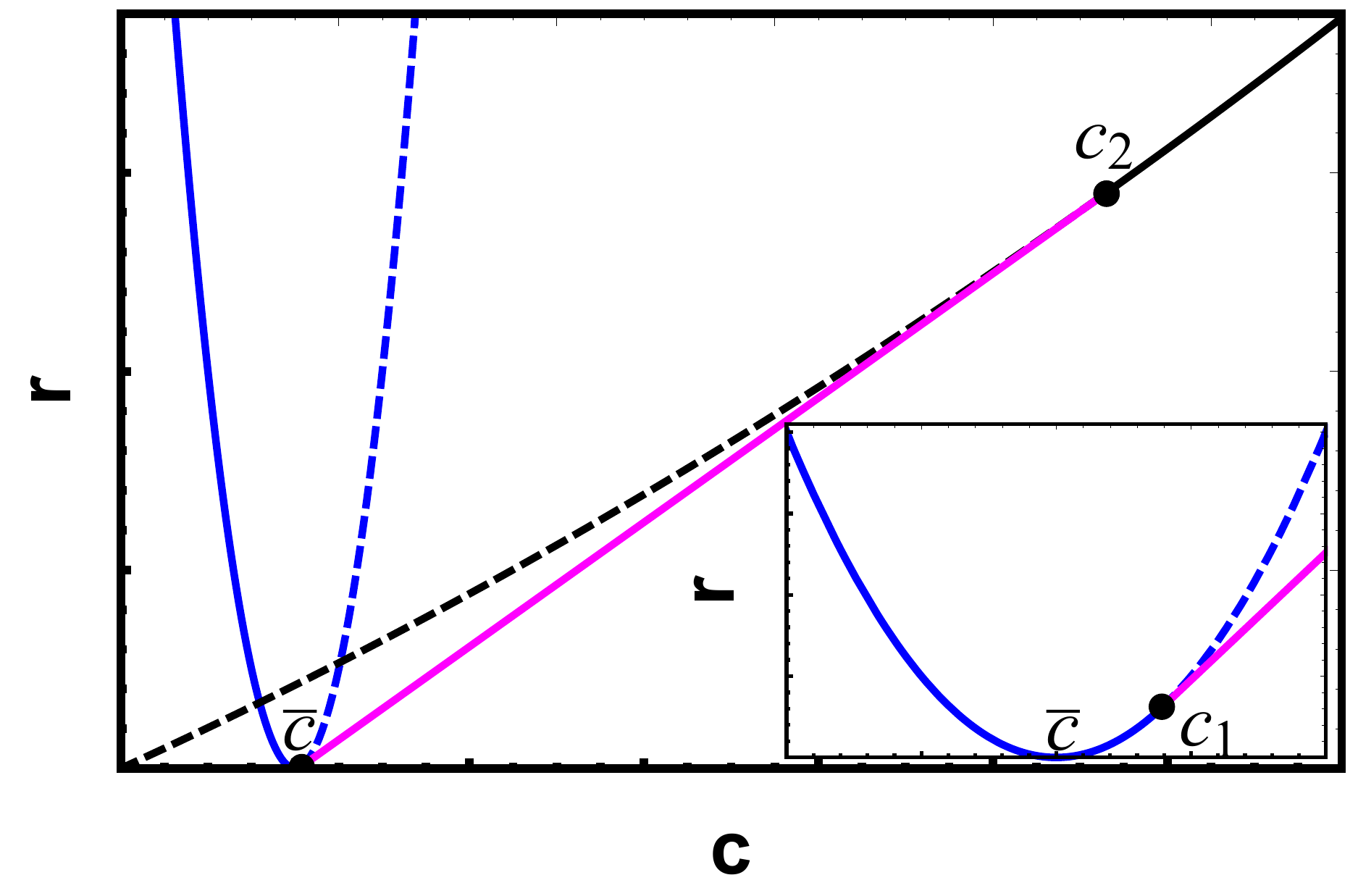}
\caption{A schematic plot of the rate function $r(c,N)$, see the text around Eqs.~(\ref{lnP})-(\ref{rBBM})}.
\label{tangentconst}
\end{figure}

The asymptotics (\ref{rgauss}) and (\ref{rBBM}) of the rate function $r(c,N)$ are schematically depicted in Fig.~\ref{tangentconst}. Both are parabolas, but the parabola (\ref{rgauss}) is much steeper: it strongly depends on $N\gg 1$, whereas the parabola (\ref{rBBM}) does not depend on $N$ in the leading order. The behavior of $r(c,N)$ in the intermediate region $c-\bar{c}\sim \bar{c}$ is presently unknown. An \emph{upper bound} for $r(c,N)$ [which gives a lower bound for the probability distribution $P(c, \Delta t, N)$] can be obtained via a tangent construction: we draw a straight line, tangent to both parabolas, see Fig.~\ref{tangentconst}. Let us denote the tangency points by $c_1$ and $c_2$, so that $c_1<c_2$. Since $N\gg 1$, the tangency point $c_1$ is very close to $\bar{c}$.  A point on the tangent line corresponds to a front history where the front  moves with velocity $c_1$ during the time $\Delta t_1$, and with velocity $c_2$ during the time $\Delta t_2$, where
$$
\Delta t_1 =\Delta t \frac{c_2-c}{c_2-c_1}\,,\quad \text{and}\quad \Delta t_2 =\Delta t \frac{c-c_1}{c_2-c_1}\,,
$$
so that $\Delta t_1 +\Delta t_2 =\Delta t$.

Now we can look into the role of large deviations or, more precisely, of unusually fast particles outrunning the front, in the fluctuations of the empirical velocity of the front. For that purpose let us define the \emph{apparent} diffusion constant of the front during the time interval $\Delta t$ in the following way:
\begin{equation}\label{Dstar}
D_*(\Delta t,N) = \frac{1}{2}\sigma^2_c(\Delta t) \Delta t \,,
\end{equation}
where
\begin{equation}\label{variance}
\sigma^2_c (\Delta t,N) =  \int_{-\infty}^{\infty} dc \left(c-\bar{c}\right)^2 P(c, \Delta t, N)
\end{equation}
is the variance  of the empirical velocity of the front $c=\Delta X/\Delta t$.

When the integral in Eq.~(\ref{variance}) is dominated by the Gaussian asymptotic (\ref{rgauss}),  the variance $\sigma^2_c$ is inversely proportional to $\Delta t$ and, by virtue of Eq.~(\ref{Dstar}),  the apparent diffusion constant $D_*$ becomes independent of $\Delta t$, and equal to the diffusion constant of the front $D_f$. For this to happen,
$\Delta t$ must be sufficiently large, so that the effective integration length in Eq.~(\ref{variance}) is within the applicability of the perturbation theory of subsection~\ref{diffusion}. This immediately leads to the strong inequality $\Delta t \gg N$.

At intermediate times $1\ll \Delta t \ll N$, the variance $\sigma_c^2$ is dominated by the positive large-deviation tail of the distribution $P(c, \Delta t,N)$, which is almost independent of $N$. This fact has two important consequences. First, the apparent diffusion constant $D_*$ in Eq.~(\ref{Dstar}) does depend on time. Therefore, the fluctuations of the empirical velocity of the front in this regime are non-diffusive. Second, these fluctuations are very large (almost independent of $N$).  Strikingly, the larger $N$ is, the longer is the transient regime where the fluctuations of the front position are anomalously large and non-diffusive.

Finally, the positive large-deviation tail of the distribution $P(c, \Delta t,N)$ is insensitive (in the leading, zeroth order in $1/N$) to whether the front is strongly or weakly pushed. Therefore, the same intermediate-time anomalies should be also observed for weakly-pushed fronts.

\section{Simulations}
\label{simulations}

\subsection{General}

To test our theoretical predictions, we performed extensive Monte Carlo simulations of the stochastic reacting lattice gas model described above. We put particles on a one-dimensional lattice with unit spacing such that every lattice site can be occupied by any number of particles. The length  $L$ of the simulated system depended on the parameters. It was chosen to allow for reliable measurements of the front displacement for different time intervals in the steady-front regime and varied from $L=2000$ to $L=8000$. The initial particle density corresponded to the theoretical deterministic profile (\ref{ansol}).

Because of the single-step character of each of the three processes, $A\to 0$, $2A \to 3A$ and $3A \to 2A$, there is an effective birth-death process on each site  and random walk along the lattice. That is, a particle can be born with the birth rate $\chi=\alpha + \beta(n_i-1)/2$, die with the death rate  $\mu= \sigma(n_i-1)(n_i-2)/6$, or hop to any of the two neighboring sites with the hopping rate $D_0$.  Since the number of particles $n_i$ on each site is different, the rates change from site to site.

To perform Monte Carlo simulations, we employed the standard Gillespie algorithm \cite{Gillespie}. First, a site was chosen with probability proportional to the overall \emph{activity} on that site, which is computed as the number of particles on the site times the sum of all the rates there. Notice that, when the hopping rate $D_0$ is much larger than the reaction rates, the
sum of all the rates is dominated by $D_0$. Therefore, for a constant hopping rate, the activity-based sampling can be simplified: a site can be chosen with the probability proportional just to the number of particles on that site \cite{densitydependent}.

Once a site is chosen, a particle on that site is chosen at random, and it performs one of the three processes, with probabilities proportional to the rates: $p_{\text{hopping}} = D_0/(D_0+\chi+\mu)$, $p_{\text{birth}} = \chi/(D_0+\chi+\mu)$, and $p_{\text{death}} = \mu/(D_0+\chi+\mu)$. If a hopping process is chosen, a particle jumps to the right and to the left with an equal probability of $1/2$. After every single-particle event, the time is advanced by
$$
\delta t= \frac{1}{M (D_0+\chi+\mu)}\,,
$$
where $M$ is the total number of particles in the system. A no-flux boundary condition was implemented at $i=0$, and we made sure that the position of the rightmost particle is always smaller than the chosen system length $L$.

\begin{figure}[ht]
\includegraphics[width=0.42\textwidth,clip=]{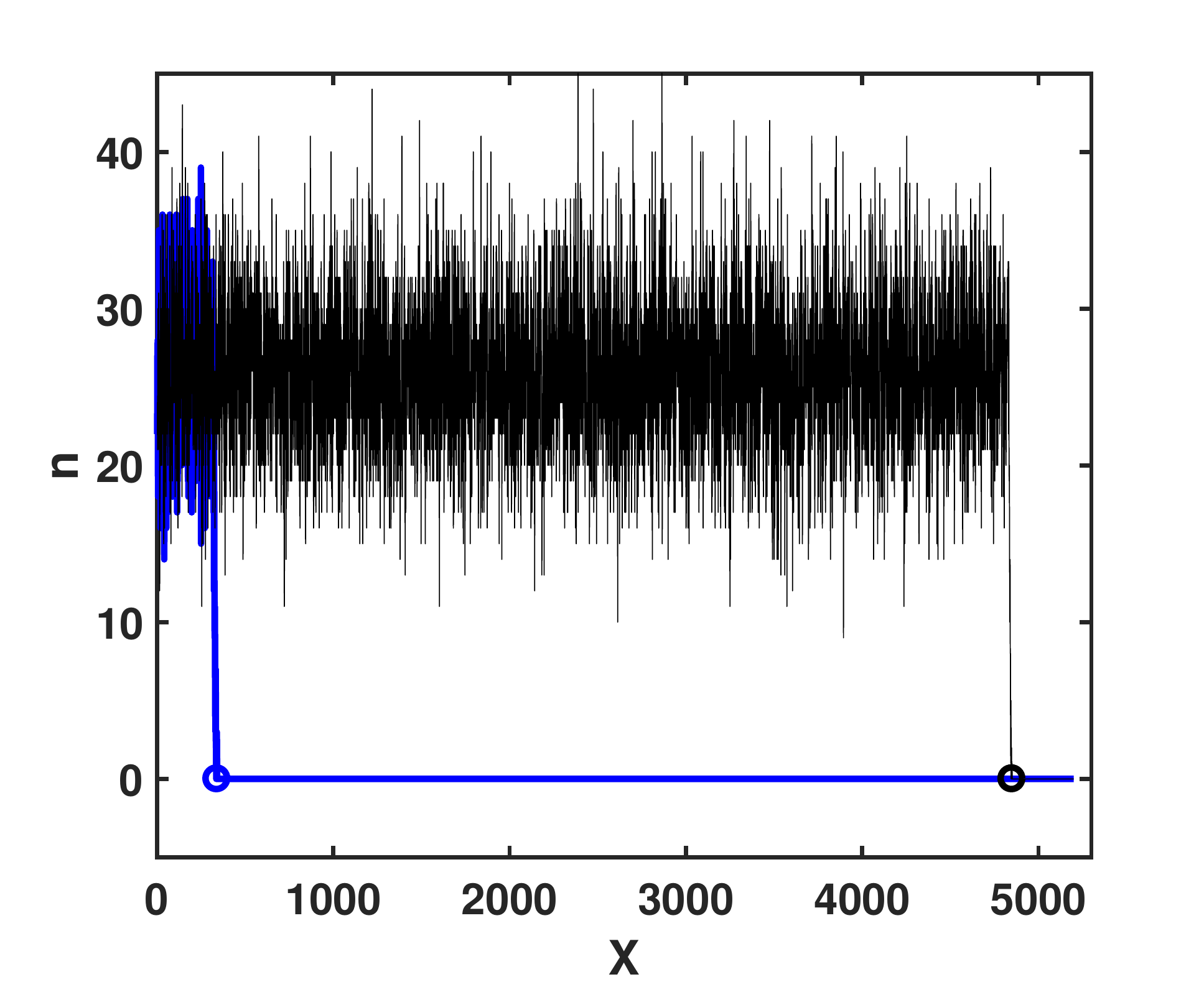}
\caption{Measuring the empirical velocity of the front.  Shown is the number density of the particles versus the integer coordinate $X=i$. The blue line corresponds to the reference time $t_0$, the black line corresponds to time $t_f$. The positions of the rightmost particles at these two time moments are denoted by circles. The parameters are: $\alpha=1$, $\beta = 4/15$, $\sigma = 9/225$,  $D_0 = 50$, $t_0=20$, $t_f=440$, and $L=5200$. For these parameters $K=10$, $\gamma = 3/2$, and $N=50$.}
\label{fig:frontprofile}
\end{figure}

There are several practical methods of measuring the position and empirical velocity of stochastic fronts. One method~\cite{KM} is to fit the instantaneous stochastic front to the theoretical deterministic front solution~(\ref{ansol}), as shown in Fig.~\ref{fig:frontprofile2}. The only adjustable parameter here is the front position. In this work we chose a more straightforward method by tracking the (integer) position of the rightmost particle. \cite{variant}.
Denoting this position by $X_f$ (here we will use $X$ instead of $i$ for convenience), we followed $X_f$  in time, and computed the empirical velocity of the front on the time interval $(t_0, t_f)$ as $\Delta X/\Delta t$, where $\Delta X=X_f - X_0$, $X_0$ is the position of the front at time $t_0$ (after the front reached a steady state), $X_f$ is the position at time $t_f>t_0$, and $\Delta t=t_f-t_0$. An example of such a measurement is shown in Fig.~\ref{fig:frontprofile}. One can see the stochastic front in two positions: at $t=t_0=20$ and at $t=t_f=440$. The positions of the rightmost particles are denoted by circles. In this particular simulation the empirical velocity of the front is $2.1477$, which is smaller than the theoretical deterministic value $c_0=2.1613$.

The two methods of measuring the front position can give very different results at times shorter than, or comparable with, the relaxation time of the front, which is $O(1)$ in our rescaled units. At long times, however,  that we are interested in in this work, the results should be very close.
We compared the two methods for $N=25$ and $N=50$ and found that they indeed produced almost identical results. After averaging over 600 simulations, the difference in the front velocity, obtained by the two methods in these two cases, was less than $0.1$ percent, and the difference in the results for the front diffusion coefficient was about one percent.

We performed a total of about $15 000$ simulations with different parameters.
In each set of simulations for the same parameters we computed the ensemble-average $\bar{c}$ and the standard deviation  $\sigma_c$ of the empirical velocity of the front for different values of the time difference $\Delta t = t_f-t_0$. Then we analyzed  the dependence of these two quantities on $N$ and on $\Delta t$. As the computational cost of these simulations scales as $N^3$, we could reach only a limited range of $N$ which, nevertheless, was sufficient to give a strong evidence in favor of our theory, as we will see shortly.

\subsection{Simulations vs. theory}
\label{simvstheory}

\begin{figure}[ht]
\includegraphics[width=0.38\textwidth,clip=]{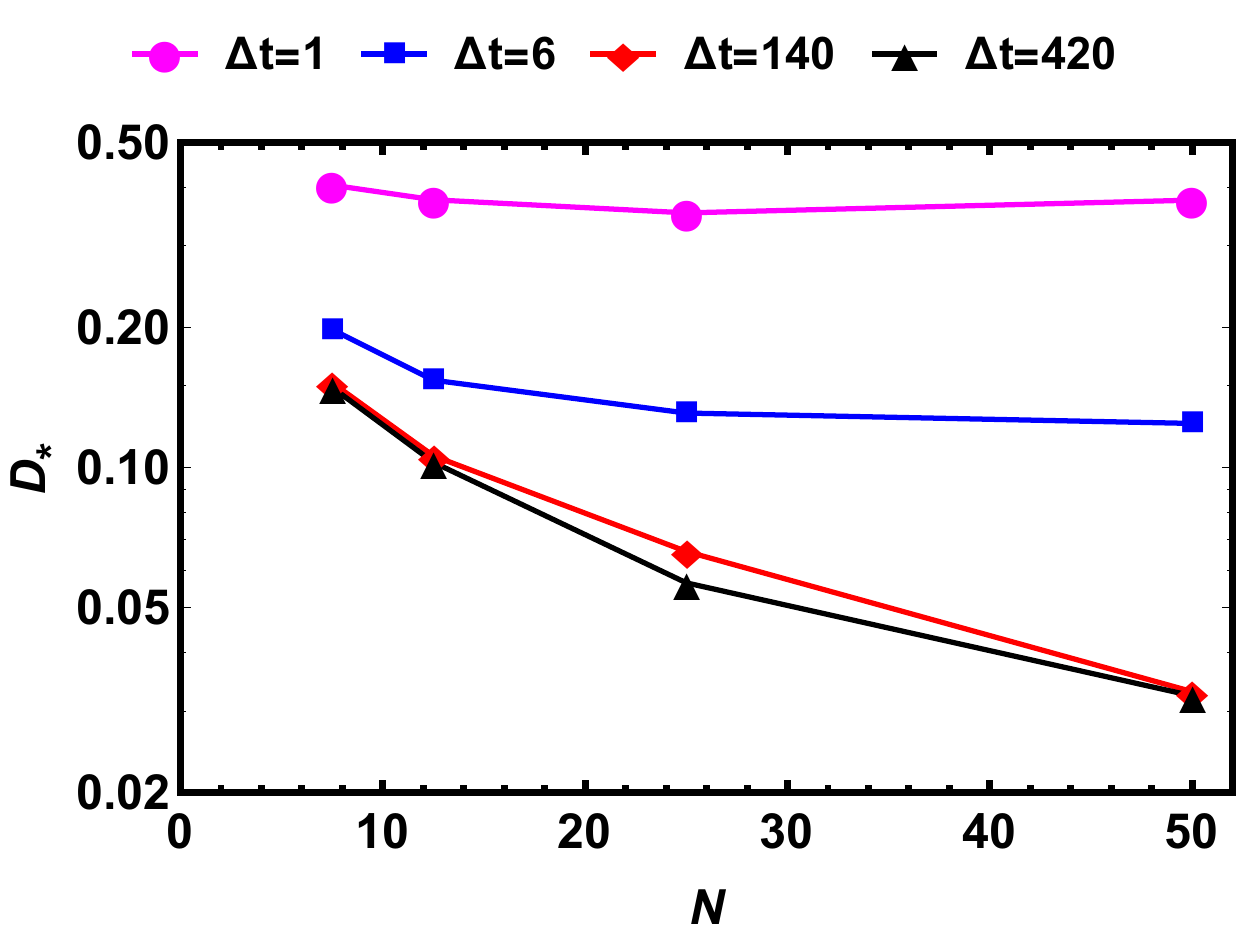}
\caption{The simulated apparent diffusion constant $D_*(\Delta t,N)$ versus $N$ for different values of $\Delta t$. A log-scale is used for the vertical axis.}
\label{fig1}
\end{figure}

\begin{figure}[ht]
\includegraphics[width=0.38\textwidth,clip=]{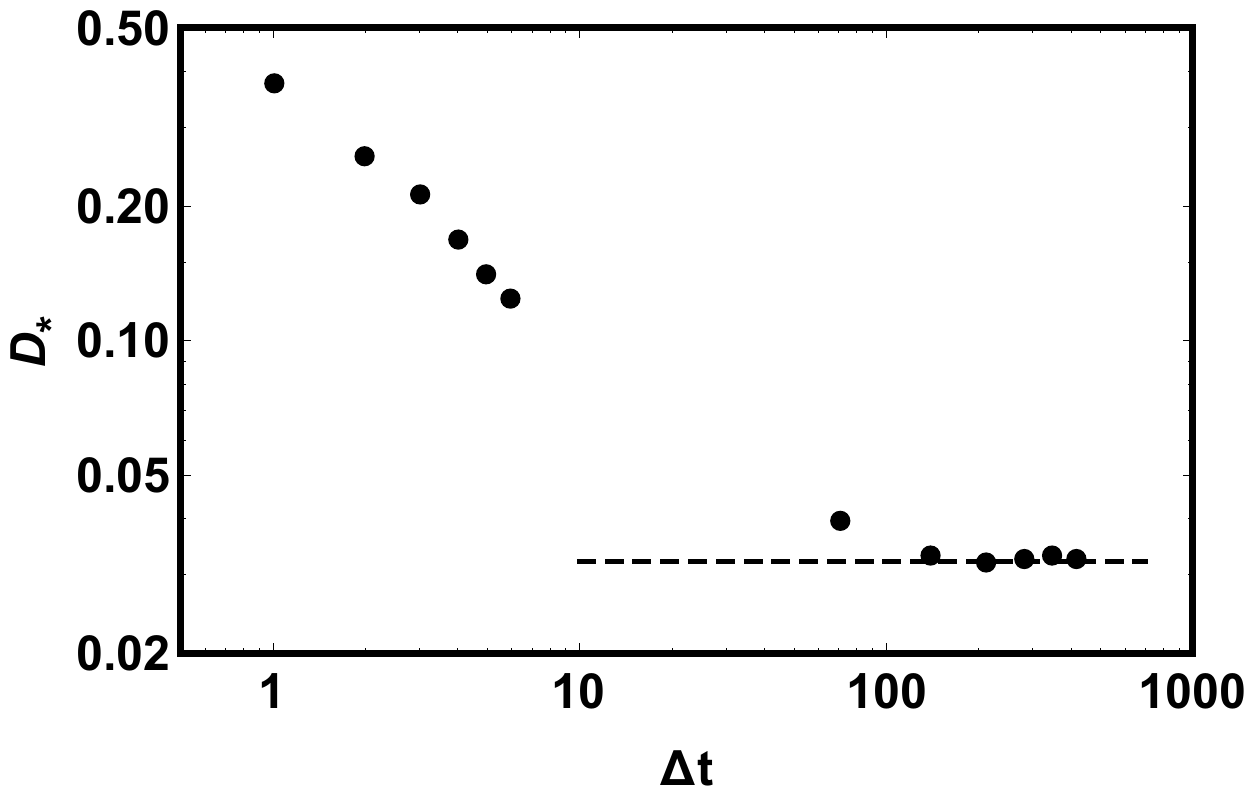}
\caption{The simulated apparent diffusion constant $D_*(\Delta t,N)$ versus $\Delta t$ for $N=50$. The dashed horizontal line shows our theoretical prediction: Eqs.~(\ref{DF070}) and (\ref{DF130}) with $k=8$. The log scale is used for both axes.}
\label{fig2}
\end{figure}

\begin{figure}[ht]
\includegraphics[width=0.38\textwidth,clip=]{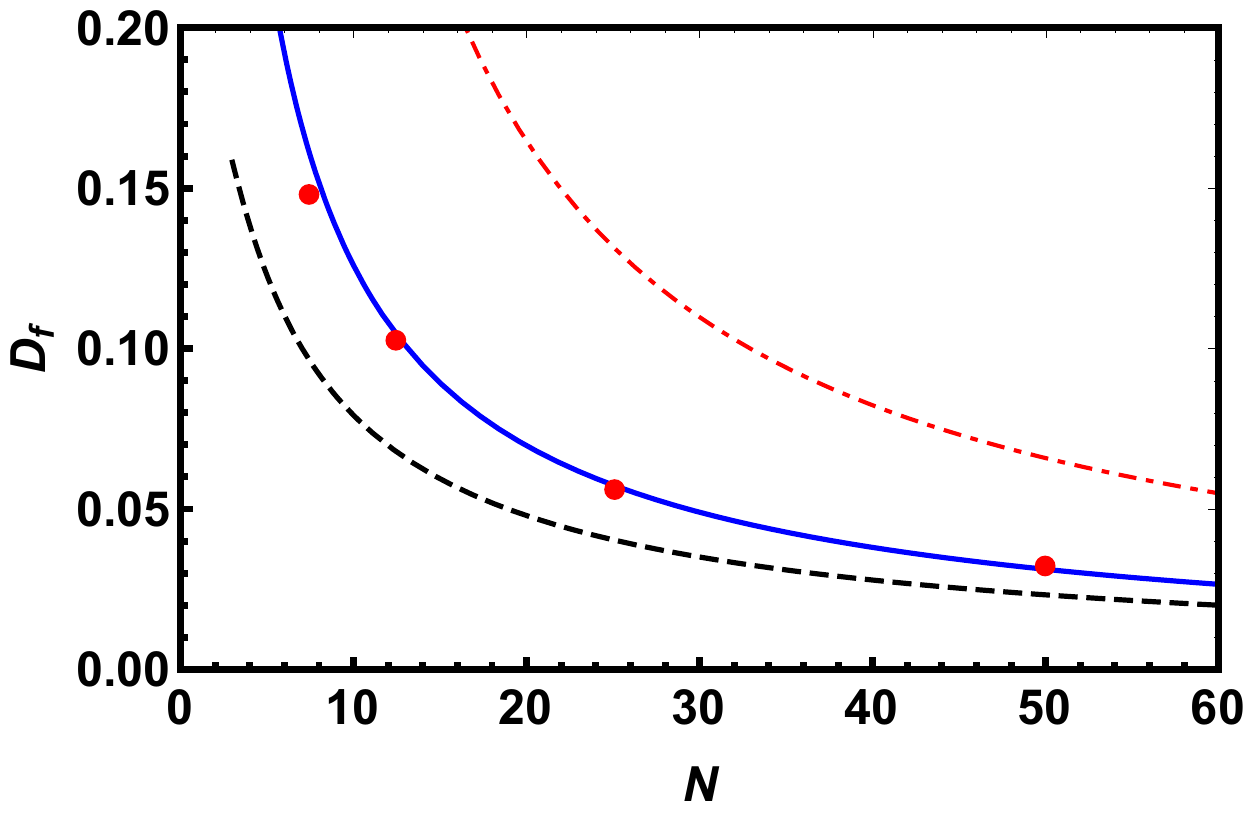}
\caption{The diffusion constant of the front, $D_f$, versus $N$. Shown are i) simulated points (the circles); ii) predictions of Eqs.~(\ref{DF070}) and (\ref{DF130}) with $k=8$ (the blue solid line), iii)  predictions of Eqs.~(\ref{DF070}) and (\ref{DF130}) with $k=1$ (the black dashed line), and iv) prediction of Eq.~(\ref{DF010}) (the red dash-dotted line). The error bars of the simulated points are not shown because they are smaller than the size of the circles. The importance of the non-perturbative $N^{-1-\nu}$ correction is clearly seen.}
\label{fig3}
\end{figure}

Figures~\ref{fig1}-\ref{fig3} summarize our simulation results on the dependence of  the fluctuations of the empirical velocity of the front on the duration of the  time interval $\Delta t$. Figure~\ref{fig1} shows the measured dependence of the apparent diffusion constant $D_*(\Delta t,N)$ from Eq.~(\ref{Dstar}) on $N$ for different values of $\Delta t$. As one can see, the $N$-dependence is strongly affected by the choice of $\Delta t$ until sufficiently large values of $\Delta t$ are reached, when $D_*$ becomes time-independent and approaches $D_f$.

The dependence of $D_*(\Delta t,N)$ on $\Delta t$ at fixed $N=50$ is shown in Fig.~\ref{fig2}. One can see that $D_*$ ceases to depend on $\Delta t$, and approaches its asymptotic value $D_f$, only when $\Delta t$ is a few times larger than $N$. This agrees with our predicted condition $\Delta t \gg N$. This feature is in agreement with our prediction that, at $\Delta t \gg N$, the lower moments of the distribution $P(c, \Delta t, N)$ are determined by the typical, Gaussian fluctuations, and the contribution of the few fast particles, outrunning the front, is suppressed. Going back to Fig.~\ref{fig1}, we see that, for $\Delta t \lesssim N$, when large positive deviations of $c$ contribute to the apparent diffusion constant $D_*(\Delta t,N)$, the dependence of $D_*(\Delta t,N)$ on $N$ is much weaker than for $\Delta t \gg N$. This supports our argument that the rate function $r(c,N)$ is almost independent of $N$ at $c-\bar{c} \gg \bar{c}$, see Eq.~(\ref{rBBM}).

Figure~\ref{fig3} focuses on the $N$-dependence of the true, long-time diffusion constant of the front $D_f$. The simulation results are shown by the filled circles. The solid line shows the prediction of Eqs.~(\ref{DF070}) and (\ref{DF130}) with $k=8$, showing a very good agreement. For comparison, the red dash-dotted line shows the leading-order prediction, Eq.~(\ref{DF010}), which ignores the integration cutoff.  Also, the black dashed line shows the prediction of  Eqs.~(\ref{DF070}) and (\ref{DF130}) with $k=1$. Here the agreement is not as good, but still much better than with the leading-order expression~(\ref{DF010}).

\vspace{0.5 cm}

\section{Summary and Discussion}
\label{summary}

We argued here, based on our theory and Monte-Carlo simulations, that fluctuations of the empirical velocity of pushed fronts can be described in terms of effective diffusion of the front only on anomalously long time intervals, $\Delta t \gg \tau_N$, where $\tau_N$ scales as $N$, the characteristic number of particles in the transition region of the front. For $\Delta t \lesssim \tau_N$ the fluctuations of the empirical velocity of the front are very large (almost independent of $N$) and non-diffusive. This prediction is striking and counterintuitive. Indeed, for macroscopic fronts, the diffusion stage of the front propagation may never be reachable, and the front fluctuations remain very large, and do not vanish ``in the thermodynamic limit" $N \to \infty$.
This anomaly is caused by a very few particles which outrun the main front, branch, reconnect with the front, \textit{etc}. This regime requires a microscopic model for its description and cannot be described by the stochastic PDE (\ref{2}).

A long non-diffusive transient should occur for weakly pushed fronts as well. The anomalously large duration of the non-diffusive transient, when $\tau_N$ scales as a positive power of $N$,  is unique for the pushed fronts. Indeed, for the pulled fronts, the large deviation form (\ref{lnP}) holds as well \cite{Derrida06,MSfisher,MSV}. The rate function $r(c,N)$ in the region of typical fluctuations behaves as \cite{Derrida06}
\begin{equation}\label{rpulledgauss}
r(c,N) \simeq \frac{\left(c-\bar{c}\right)^2}{4 D_f} \sim \ln^3 N \left(c-\bar{c}\right)^2\,,
\end{equation}
and a similar argument leads to a much shorter transient, $\tau_N \sim \ln^3 N$, characterized by large fluctuations and a non-diffusive behavior of the front. For fronts propagating into a metastable state, the rate function $r(c,N)=N \phi(c)$ is proportional to $N$ for all $c$ \cite{MSK}. As a result, the non-diffusive transient is quite short: its duration is of the same order of magnitude as the relaxation time of the deterministic front to its asymptotic shape, and is therefore independent of $N$.

In the asymptotic regime $\Delta t \gg \tau_N$ the velocity fluctuations of the strongly pushed front are small and diffusive. Here a more careful treatment of a few first particles in the leading edge of the front yields a sizable non-perturbative negative correction to the diffusion coefficient of the front. This correction becomes crucial close to the strong-weak transition and leads to a logarithmic correction to the $1/N$ scaling of $D_f$ at the transition point.  By contrast, for the fronts propagating into an empty region of space which is \emph{metastable} deterministically, a similar correction to $D_f$ is less significant. As one can show, it is much smaller than $1/N^2$.

Finally, since the function $A(N)$, given by Eq.~(\ref{DF160}), is an analytic function of $\nu$ at the transition point $\nu=0$ between the strongly and weakly pushed fronts, we conjecture that Eq.~(\ref{DF160}) remains valid for negative $\nu$ as well, once
$|\nu|\ll 1$. That is, we expect that $D_f$ scales as $N^{-1-\nu}$ for the weakly pushed fronts close to the transition point.

\section*{Acknowledgments}
We are very grateful to Bernard Derrida for useful discussions. BM was supported by the Israel Science Foundation (Grant No. 807/16). P.S.'s work is supported by the project ``High Field Initiative" (CZ.02.1.01/0.0/0.0/15\_003/0000449) of the European Regional Development Fund.


\begin{thebibliography}{99}
\bibitem{disparity} Once the parameter $N = K\sqrt{D_0/2\alpha} \gg 1$, the strong inequality  $\beta \gg \sigma$ (which is equivalent to $K\gg 1$) is unnecessary.
\bibitem{densitydependent} This method would not work, however, if the hopping rate  on a given site would depend on the number of particles on that site.
\bibitem{variant} A variant of this method \cite{KLS}, used to measure the mean front velocity,  is to track the position of the rightmost particle at multiple intermediate times, and to fit the resulting time dependence by a straight line. The slope of the straight line is identified with the mean front velocity.
\end{thebibliography}
\end{document}